\documentclass[twocolumn,secnumarabic,amssymb,amsmath,superscriptaddress,aps,pra,longbibliography]{revtex4-2}

\usepackage{hyperref}
\usepackage{graphicx}
\usepackage{amsfonts}
\usepackage{amssymb}
\usepackage{bm}
\usepackage{color}
\usepackage{xspace}
\usepackage{cancel}
\usepackage{braket}
\usepackage{chemformula}
\usepackage{array}
\usepackage{ifthen}
\usepackage{cleveref}

\newcommand*\figurewidth{0.96\columnwidth}

\newcommand\smo{\ensuremath{\mathfrak{S}}}
\newcommand\sm[1]{\ensuremath{\smo\!\left(#1\right)}}
\newcommand\tX{\ensuremath{\text{\sc x}}}
\newcommand\tY{\ensuremath{\text{\sc y}}}
\newcommand\tZ{\ensuremath{\text{\sc z}}}
\newcommand\tU{\ensuremath{\text{\sc u}}}
\DeclareMathOperator{\var}{Var}
\DeclareMathOperator{\erf}{erf}
\allowdisplaybreaks

\begin{document}

\title{Statistical theory of cumulant mapping in an imperfect apparatus}

\author{S.\ Patchkovskii}
\email{serguei.patchkovskii@mbi-berlin.de}
\affiliation{Max-Born-Institute, Max-Born-Str. 2A, 12489 Berlin, Germany} 
\author{J.\ Mikosch}
\email{mikosch@uni-kassel.de}
\affiliation{Institut f\"ur Physik, Universit\"at Kassel, Heinrich-Plett-Str. 40, 34132 Kassel, Germany}

\date{\today}

\begin{abstract}
Cumulant mapping has been recently suggested [Frasinski, Phys. Chem. Chem.
Phys. \textbf{24}, 207767 (2022)] as an efficient approach to observing
multi-particle fragmentation pathways, while bypassing the restrictions of the
usual coincidence-measurement approach. We present a formal analysis of the
cumulant-mapping technique in the presence of moderate external noise, which
induces spurious correlations between the fragments. Suppression of
false-cumulant signal may impose severe restrictions on the stability of the
experimental setup and/or the permissible average event rate, which increase
with the cumulant order. 
We demonstrate that cumulant mapping in an imperfect apparatus
remains competitive for dominant processes and for pathways with a 
background-free marker fragment.
We further show that the false-cumulant contributions increase faster than
linearly with the average event rate, providing a simple test for 
the experimental data analysis.
\end{abstract}
\maketitle

\section{Introduction\label{sec:introduction}}

Covariance mapping\cite{Frasinski89a,Frasinski2016:jphysb} is an ingenious statistical analysis of
multi-particle fragmentation experiments, which allows fragment correlations to
be established without a restriction of a single event per measurement shot,
and with only moderate requirements for the fragment detection efficiency.
Covariance-based measurements can often reach statistical significance in a
fraction of time\cite{boguslavskiy2012:sci,allum2021:jpcl}, which would have been required for
the more traditional\cite{Brehm67a} coincidence-based approaches. Covariance
mapping has enabled remarkable advances in experiments where isolated molecules
are broken up, such as X-ray and strong-field induced dissociation and Coulomb
Explosion Imaging of small
molecules\cite{vallance2021:jpca,crane2022:pccp,cooper2021:jpca}, in particular
with ultra-bright Free Electron Lasers (FELs)
\cite{allum2021:faraday,allum2022:comchem,mcmanus2022:pccp,walmsley2024:jphysb}
- including pump-probe studies of molecular
dynamics\cite{allum2018:jcp,unwin2023:comphys,mogol2024:physrevres}. Covariance
mapping is constantly further conceptually
refined\cite{walmsley2024:jpca,mcmanus2024:jpca} and adopted for structure
determination of more complex systems, such as isolated biomolecules via
collision-induced fragmentation\cite{driver2023:jacs}, as well as imaging of
tetracene dimers formed inside\cite{schoeder2019:strdyn} and of alkali trimers
produced on the surface of\cite{kranabetter2024:jcp} helium nanodroplets via
femtosecond laser-induced Coulomb explosion. Recently, an extension of the
covariance mapping to multiparticle correlations, the cumulant mapping, has
been proposed\cite{Frasinski22a}. First experimental realization of the
technique have already appeared\cite{cheng2023:prl}, as well as mathematical
refinements of the statistical treatment\cite{Andersson23a,cheng2024:pra}. 

The Achilles' heel of the covariance-mapping techniques are the spurious
correlations between fragments, induced by fluctuations of the experimental
parameters, rather than due to the common origin of the fragments in a
molecular fragmentation process. Examples of such ``external'' fluctuations 
include the target density or the laser pulse fluence
and/or intensity\cite{Mikosch13a,Mikosch13b,zhaunerchyk2014:pra}. Such
fluctuations are an inevitable consequence of experimental imperfections, and
induce ``false'' covariances between all fragments, potentially swamping weak
channels of interest. If an independent, shot-by-shot measurement of the
fluctuating parameter is available\cite{li2022:comphys, dingel2022:scirep},
their effect can be reversed using ``partial'' covariances\cite{Kornilov13a}
or ``contingent'' covariances\cite{zhaunerchyk2014:pra}.
Alternatively, simultaneous detection of a sufficient number of fragmentation
channels, which become accessible as the complexity of the system being
studied increases, can be used to implement a \textit{self-correcting} partial
covariance\cite{Driver20a}, even when no explicit measurement of the
fluctuating external parameter(s) is available. 

A statistical analysis of covariance measurement in an imperfect apparatus,
placing constraints on the permissible noise level and event rates, has been
available in the literature for some time\cite{Mikosch13a,Mikosch13b}.
Unfortunately, we are not aware of a comparable investigation of the cumulant
mapping. The goal of this contribution is to fill this lacuna, using techniques
developed in Ref.~\cite{Mikosch13a} and previously applied to the analysis of
self-correcting covariances\cite{Driver20a}.

The rest of this work is organized as follows: Section~\ref{sec:montocarlo}
demonstrates the appearance of ``false'' covariances in a 3-cumulant via a
numerical experiment, Section~\ref{sec:notation} establishes notation and
recapitulates key results from Ref.~\cite{Mikosch13a}, necessary to follow the
discussion. Section~\ref{sec:results} presents the expectations and the
variances of the 2-, 3-, and 4-cumulants in an imperfect apparatus.
Section~\ref{sec:discussion} discusses the constraints imposed by these results
on the experimental parameters, in a number of typical scenarios, and briefly
examines the convergence of cumulant maps.  Finally,
section~\ref{sec:conclusions} summarized the work and offers the outlook for
future developments.

\section{The prelude: Numerical Experiment\label{sec:montocarlo}}

To illustrate the appearance of ``false'' cumulants due to noise-induced
correlations, we perform a numerical Monte-Carlo simulation, fashioned after a
realistic experimental situation. A generic triatomic molecule, ABC, breaks up
via Coulomb Explosion Imaging driven by an ultrashort, intense infrared or
X-ray laser pulse, which ionizes the molecule. The desired channel, delivering
structural information, is the complete fragmentation into A$^+$ + B$^+$ +
C$^+$. This channel is assumed to be weak -- we adopt statistical probability
of 1$\%$. The incomplete fragmentation channels A$^+$ + BC$^+$, AC$^+$ + B$^+$,
and AB$^+$ + C$^+$ are taken to be much more likely. For simplicity, and
without qualitatively affecting our conclusions, we assume that each of these
incomplete channels occurs with an equal statistical probability of 33$\%$. 

The benefit of covariance over coincidence detection is to be able to perform
an experiment in a regime where multiple break-up events occur per laser shot.
In our simulation, the number of break-up events per shot is drawn from a
noise-augmented Poisson distribution. As is more rigorously put in equations in
the following section, the noise implies that the average event rate $\nu$ of a
Poisson distribution is no longer a fixed parameter, but is instead sampled
from normal distribution with the mean $\nu_0$ and standard deviation
$\sigma\nu_0$. For simplicity, we initially assume an unrealistic detection
efficiency for fragments of $\beta$ = 100$\%$. Note that in the terminology
used by Frasinski\cite{Frasinski22a}, this numerical experiment is entirely
``noise-free'', since no undesired correlated fragmentation pathways are
present. 

We used $N=5\times10^7$ shots for each simulation, which we repeated 10 times
to obtain the average of the 3-cumulant and its standard deviation. The noise
parameter $\gamma$, the number of events per shot $\nu$, the fragmentation
channel assigned to each event, and whether the individual fragments are
detected (for $\beta$ $<$ 100$\%$) are all independently statistically
determined, via drawing random numbers from Gaussian, Poissonian, and uniform
distributions. In each simulation, we loop through the $N$ shots twice - once
to determine the average rates and a second time to calculate the 3-cumulant.
Importantly, the same pseudo-random numbers are used in the two passes of the
simulation, ensured by using the same random seed - which is otherwise randomly
generated for each simulation. Since the simulation is carried out
event-by-event, we can determine not only the overall 3-cumulant, but also the
``true'' 3-cumulant, which results from the desired three-body fragmentation
channel.

\begin{figure}[!tbhp]
  \includegraphics[trim=0pt 0pt 0pt 0pt,clip,width=\figurewidth]{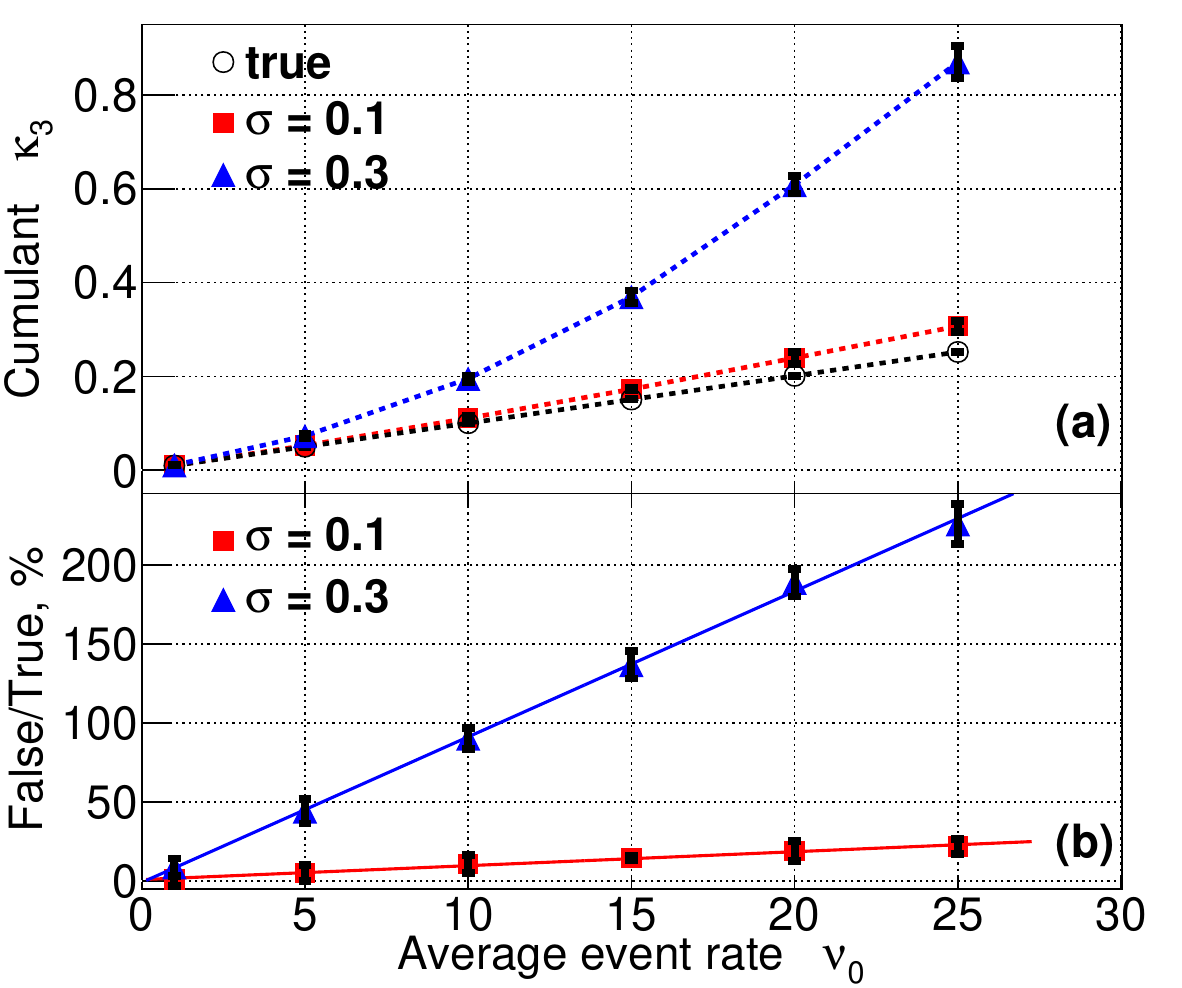}
  \caption{\label{fig:demo} 
Results of a numerical Monte-Carlo simulation, demonstrating how external
noise, such as shot-to-shot fluctuations of laser power or pulse duration,
leads to very significant deviations of the cumulant observable from the true
value. The situation in the simulation represents a minor process with strong,
but uncorrelated background - the three-body breakup of a molecule ABC into A + B + C
(1$\%$ probability), with background from A + BC, AC + B, and AB + C (33$\%$
probability each).}
\end{figure}

In Fig.\ref{fig:demo}(a), the numerically measured 3-cumulant $\kappa_3$ is
plotted as a function of the average event rate $\nu_0$, for standard
deviations $\sigma$ of the noise distribution of 0.1 (10$\%$ noise / red
squares) and 0.3 (30$\%$ noise / blue triangles). The ``true'' 3-cumulant is
plotted as black circles. Error bars represent 10 standard deviations, and are
in most cases below the size of the markers. Dashed lines connect the
simulation results for visual guidance. The ``true'' 3-cumulant, stemming from the
desired triple-fragmentation events alone, is found to increase linearly with
average event rate $\nu_0$. At the same time, the relative statistical error is
found to decrease with $\nu_0$. This observation represents the ``covariance
advantage''. In the absence of noise, the 3-cumulant $\kappa_3$ is a strongly
increased correlation signal, as compared to the coincidence detection regime
which requires $\nu_0 \ll 1$ to avoid false coincidences. 

However, Fig.\ref{fig:demo}(a) also shows how the ``true'' 3-cumulant of the
desired three-body fragmentation channel is increasingly ``polluted'' by the
coincidental two-body fragmentation channels with increasing noise. For larger
$\sigma$, $\kappa_3$ deviates increasingly from the linear dependence on
$\nu_0$, eventually coming to dominate the signal. Also the relative error is
strongly increased, as can be seen from the appearance of error bars that are
larger than the marker size. In Fig. \ref{fig:demo}(b) the ratio of false to
true $\kappa_3$ extracted from the data in panel (a) is displayed, along with
linear fits. This ratio, marking the systematic error to the desired ``true''
3-cumulant, is found to increase linearly with average event rate $\nu_0$, and
the slope is strongly increased for $\sigma$ = 0.3 as compared to $\sigma$ =
0.1. 

The simulation was repeated for a reduced detection efficiency, $\beta=0.5$
(50$\%$ fragment detection probability). We observe that 3-cumulant is reduced
by $\beta^3$ and that its relative error is increased. The ratio of false to
true 3-cumulant, however, stays the same, albeit with a larger statistical
error.

The results of the numerical experiment as shown in Fig.\ref{fig:demo} caution
that in covariance detection the average event rate $\nu_0$ has to be kept
below a limit that is given by a systematic error to the n-cumulant that is
deemed tolerable. This issue arises when external noise is present, such as
statistical fluctuations of the laser power or pulse duration - an omnipresent
situation in real experiments, in particular with FELs. The systematic error by
``false'' covariances strongly increases for increasing noise, which can
restrict the beneficial range of an average event rates $\nu_0$ severely. Thus
motivated, we are now ready to turn to a more formal analysis.

\section{Notation\label{sec:notation}}

We consider fragmentation of independent particles, possibly of different
kinds, triggered by an external event (a ``shot'', such as a light pulse).
Each primitive event produces fragments of interest ($1=\tX$, $2=\tY$, $3=\tZ$,
$4=\tU$, etc.), plus possibly additional fragments of no concern to us.  A
fragment label is understood to include not only the intrinsic nature of the
particle, but also any relevant bin label assigned in the data analysis. For
example, if a water molecule \ch{H2O} is Coulomb-exploded in the experiment, we
could assign label $\tX$ to \ch{O+} ions, label $\tY$ to \ch{H+} ions with the
kinetic energy below $1$~eV, and label $\tZ$ to \ch{H+} ions with the kinetic
energy above $4$~eV.

In each shot, the probability $P\left(n\right)$ of $n$ independent
fragmentation events occurring follows noise-augmented Poisson
distribution\cite{Mikosch13a}:
\begin{align}
  P_\nu\left(n\right) &= \frac{1}{n!} \nu^n \exp\left(-\nu\right), & \label{eqn:notation:pnu}
\end{align}
where the primitive event rate $\nu=\nu_0\gamma$ is sampled from a normal
distribution of the relative width $\sigma$, centered about the average
rate $\nu_0$:
\begin{align}
  P_\sigma\left(\gamma\right) &= \frac{1}{\sqrt{2\pi} \sigma}
  \exp\left(-\frac{\left(\gamma-1\right)^2}{2\sigma^2}\right). & \label{eqn:notation:psigma}
\end{align}
We assume that $\sigma$ is sufficiently small, so that
$P_\sigma\left(\gamma<0\right)$, and therefore the probability of (unphysical)
negative event rates, is negligible.

Possible outcomes of each elementary event (e.g. detection of particles $X$ and
$U$) are assumed to be mutually-exclusive. Each elementary outcome is described
by the corresponding probability $P_i$. The probability vector of all $m$
relevant outcomes is denoted $\mathbb{P}$. Similarly, in each shot the fragment
counts are given by elements of a vector $\mathbb{N}$. The moments
$M\left(\mathbb{K}\right)$ of the probability distribution
$P\left(\mathbb{N}\right)$ are defined as:
\begin{align}
  M\left(\mathbb{K}\right) & = \sum_{n_1=0}^\infty ... \sum_{n_m=0}^\infty 
                               P\left(\mathbb{N}\right) \prod_{i=1}^{m} n_i^{k_i}. \label{eqn:notation:moment}
\end{align}

The moments $M\left(\mathbb{K}\right)$ can be conveniently evaluated using a
recursive expression, derived in\cite{Mikosch13a}:
\begin{align}
  &M\left(\mathbb{K}+\mathbb{I}_j\right) = 
     P_j \nu_0 M\left(\mathbb{K}\right) + P_j \frac{\partial}{\partial P_j} M\left(\mathbb{K}\right)
     & \nonumber \\
     & \quad
     + \nu_0^2 \sigma^2 P_j \!\sum_{l=1}^{m} P_l \!\sum_{b_l=0}^{k_l-1} \binom{k_l}{b_l} 
       M\left(\mathbb{K}+\mathbb{I}_l \left(b_l-k_l\right)\right)
  , & \label{eqn:notation:momentsrecursive} \\
  &M\left({\bf 0}\right) = 1, & \label{eqn:notation:bootstrap}
\end{align}
where $\mathbb{I}_i$ denotes a vector of length $m$, with $1$ at position $i$
and zeros at all other indices, and the norm of the distribution
$M\left({\bf0}\right)$ provides the necessary bootstrap.

At the most fundamental level, the fragmentation and detection process are
described by the probabilities of fragment formation in an elementary event and
their detection in the experimental apparatus. For example, we denote the
probability of particles $X$ and $Y$, \textit{and no other particles of
interest}, being produced by a primitive fragmentation event $\alpha_{XY}$.
The corresponding probabilities of detection, once particles are produced, are
denoted $\beta_X$ and $\beta_Y$. It is convenient to combine the parameters
$\alpha$ and $\beta$, which describe the microscopic mechanism of the
fragmentation process, into parameters more closely related to the experimental
observation. For example, for the case of the 3-particle cumulant, we define:
\begin{align}
  \gamma_{\tX\tY\tZ} &= \alpha_{\tX\tY\tZ}\beta_\tX \beta_\tY \beta_\tZ, & \label{eqn:notation:gamma}  \\
  \gamma_{\tX\tY}    &= \left(\alpha_{\tX\tY\tZ}+\alpha_{\tX\tY}\right) \beta_\tX \beta_\tY, & \nonumber \\
  \gamma_{\tX\tZ}    &= \left(\alpha_{\tX\tY\tZ}+\alpha_{\tX\tZ}\right) \beta_\tX \beta_\tZ, & \nonumber \\
  \gamma_{\tY\tZ}    &= \left(\alpha_{\tX\tY\tZ}+\alpha_{\tY\tZ}\right) \beta_\tY \beta_\tZ, & \nonumber \\
  \gamma_{\tX}       &= \left(\alpha_{\tX\tY\tZ}+\alpha_{\tX\tY}+\alpha_{\tX\tZ}+\alpha_\tX\right) \beta_\tX, & \nonumber \\
  \gamma_{\tY}       &= \left(\alpha_{\tX\tY\tZ}+\alpha_{\tX\tY}+\alpha_{\tY\tZ}+\alpha_\tY\right) \beta_\tY, & \nonumber \\
  \gamma_{\tZ}       &= \left(\alpha_{\tX\tY\tZ}+\alpha_{\tX\tZ}+\alpha_{\tY\tZ}+\alpha_\tZ\right) \beta_\tZ. & \nonumber
\end{align}
Quantity $\gamma_{\tX\tY}$ of \cref{eqn:notation:gamma}, for example, are to
be understood as the probability of a single, primitive fragmentation process
leading to particles $\tX$ and $\tY$, plus possibly any other particles, such
as $\tZ$, being detected. 

Clearly, events described by probabilities $\gamma$ are not mutually exclusive:
detection of particles $\tX$, $\tY$, and $\tZ$ implies that \textit{all other}
events in \cref{eqn:notation:gamma} have also occurred. To use these quantities
in \cref{eqn:notation:momentsrecursive}, defined in terms of
mutually-exclusive events, we combine them as follows:
\begin{align}
  p_{\tX\tY\tZ} &= \gamma_{\tX\tY\tZ}, & \label{eqn:notation:p} \\
  p_{\tX\tY}    &= \gamma_{\tX\tY} - \gamma_{\tX\tY\tZ}, & \nonumber \\
  p_{\tX\tZ}    &= \gamma_{\tX\tZ} - \gamma_{\tX\tY\tZ}, & \nonumber \\
  p_{\tY\tZ}    &= \gamma_{\tY\tZ} - \gamma_{\tX\tY\tZ}, & \nonumber \\
  p_{\tX}       &= \gamma_{\tX} - \gamma_{\tX\tY} - \gamma_{\tX\tZ} + \gamma_{\tX\tY\tZ}, & \nonumber \\
  p_{\tY}       &= \gamma_{\tY} - \gamma_{\tX\tY} - \gamma_{\tY\tZ} + \gamma_{\tX\tY\tZ}, & \nonumber \\
  p_{\tZ}       &= \gamma_{\tZ} - \gamma_{\tX\tZ} - \gamma_{\tY\tZ} + \gamma_{\tX\tY\tZ}. & \nonumber
\end{align}
From the definitions of the parameters $\gamma$, it is easy to convince oneself
that probabilities $p$ remain non-negative. They are to be understood as
\textit{exclusive} probabilities of observing particles of interest in a
primitive fragmentation event. For example, $p_{\tX\tY}$ describes probability
of particles $\tX$ and $\tY$, and \textit{no other particles of interest} being
observed on the detector. For unit detection efficiency (all $\beta=1$),
quantities $p$ and $\alpha$ coincide.

Because the fragment labels $\tX$, $\tY$, etc are entirely arbitrary, all
expressions for the expectation and variance of the cumulants must remain
symmetric with respect to their permutation. We therefore introduce the
symmetrization operator \sm{Q}, which transforms its argument $Q$ to a sum of
terms with all possible (orderless) permutations of the particle labels, with
each term appearing exactly once. Arguments which are already
permutation-symmetric remain unchanged, so that operator $\smo$ is idempotent:
$\sm{\sm{Q}}=\sm{Q}$.  For example, for the 3-cumulant and particle indices
$\tX$, $\tY$, and $\tZ$:
\begin{align}
  \sm{\gamma_{\tX\tY\tZ}}          & = \gamma_{\tX\tY\tZ}, & \label{eqn:notation:s} \\
  \sm{\gamma_{\tX}}                & = \gamma_\tX + \gamma_\tY + \gamma_\tZ, & \nonumber \\
  \sm{\gamma_{\tX\tZ}^2\gamma_\tY} &= \gamma_{\tX\tZ}^2\gamma_\tY 
       + \gamma_{\tY\tZ}^2\gamma_\tX + \gamma_{\tX\tY}^2\gamma_\tZ. & \nonumber
\end{align}
Use of the operator \sm{Q} allows for more compact, and manifestly symmetric,
expressions.

Although application of \cref{eqn:notation:momentsrecursive} is, in principle,
straightforward, for higher-order cumulants it rapidly becomes tedious. Thus,
the recursion \eqref{eqn:notation:momentsrecursive} needs to be invoked $28$
times while evaluating the 2-cumulant and its variance, 1198 times for the
3-cumulant and variance, increasing to 370577 times in the case of 4-cumulant.
It is therefore best accomplished with the help of a computer-algebra package.
We include Mathematica\cite{mathematica121} package implementing these
derivations as a supplementary material\cite{supplementary}, and only give the
final results below.

\section{Results\label{sec:results}}

Our main results are presented in the section below. The results for the
2-cumulant (the covariance) were given before\cite{Mikosch13a}, in a less
symmetric form and under somewhat more restrictive assumptions. In the absence
of noise ($\sigma=0$), for a perfect Poisson source, our results for
$n$-cumulant coincide with those of Frasinski\cite{Frasinski22a}, provided that
only the 1- and $n$-particle fragmentation pathways are considered.

\subsection{2-cumulant (covariance)\label{subsec:2-cumulant}}

The two fragments of interest are $\tX$ and $\tY$. The covariance is sampled 
from a distribution with the mean $\kappa_2$ and variance
$\var\kappa_2$:
\begin{align}
  \kappa_2 &= \nu_0 \gamma_{\tX\tY} + \nu_0^2\sigma^2 \gamma_\tX\gamma_\tY
     ,& \label{eqn:results:kappa2} \\
  \var\kappa_2
     &= \nu_0            \gamma _{\tX\tY}
      + \nu_0^2          \left(\gamma_\tX \gamma_\tY + \gamma_{\tX\tY}^2\right) & \nonumber \\
     &+ \nu_0^2 \sigma^2 \left(\sm{\gamma_\tX \gamma_{\tX\tY}} + \gamma_\tX\gamma_\tY + 2\gamma_{\tX\tY}^2\right) & \nonumber \\
     &+ \nu_0^3 \sigma^2 \gamma_\tX \gamma_\tY \left(\sm{\gamma_\tX} + 2 \gamma_{\tX\tY}\right) & \nonumber \\
     &+ \nu_0^4 \sigma^4 \left(2 \gamma_\tX^2 \gamma_\tY^2\right)
     .& \label{eqn:results:varkappa2}
\end{align}

\subsection{3-cumulant\label{subsec:3-cumulant}}

The three fragments of interest are $\tX$, $\tY$, and $\tZ$. The 3-cumulant
$\kappa_3$ and its variance are given by:
\begin{align}
  \kappa_3 &= \nu_0\gamma_{\tX\tY\tZ} + \nu_0^2\sigma^2 \sm{\gamma_{\tX\tZ}\gamma_\tY}
     ,& \label{eqn:results:kappa3} \\
  \var\kappa_3
     &= \nu_0 \gamma_{\tX\tY\tZ} 
      + \nu_0^2 d_{20} 
      + \nu_0^3 d_{30} & \nonumber \\
     &+ \nu_0^2 \sigma^2 d_{22} 
      + \nu_0^3 \sigma^2 d_{32} 
      + \nu_0^4 \sigma^2 d_{42} & \nonumber \\
     &+ \nu_0^4 \sigma^4 d_{44}
      + \nu_0^5 \sigma^4 d_{54}
      + \nu_0^6 \sigma^6 d_{66}
     .& \label{eqn:results:varkappa3}
\end{align}
The subscripts of the coefficients $d_{ab}$ refer to the powers of $\nu_0$ and $\sigma$ in
the corresponding polynomial term $\nu_0^a\sigma^b$. The coefficients are given by:
\begin{align}
  d_{20} &=
      3\gamma_{{\tX\tY\tZ}}^2
    + 4\gamma_{{\tX\tY\tZ}} \sm{\gamma_{{\tX\tY}}} & \nonumber \\
   &+ 2\sm{\gamma_{{\tX\tY}} \gamma_{{\tX\tZ}}}
    +  \sm{\gamma_{{\tX\tZ}} \gamma_\tY}
     ,& \label{eqn:results:d20} \\
  d_{30} &=
      8\gamma_{{\tX\tY}} \gamma_{{\tX\tZ}} \gamma_{{\tY\tZ}}
    + 2\sm{\gamma_{{\tX\tZ}}^2 \gamma_\tY}
    +  \gamma_\tX \gamma_\tY \gamma_\tZ
     ,& \label{eqn:results:d30} \\
  d_{22} &=
      4\gamma_{{\tX\tY\tZ}}^2
    + 2\gamma_{{\tX\tY\tZ}} \left(\sm{\gamma_\tX}+2 \sm{\gamma_{{\tX\tY}}}\right) & \nonumber \\
   &+ 2\sm{\gamma_{{\tX\tY}} \gamma_{{\tX\tZ}}}
    +  \sm{\gamma_{{\tX\tZ}} \gamma_\tY}
     ,& \label{eqn:results:d22} \\
  d_{32} &=
     24\gamma_{{\tX\tY}} \gamma_{{\tX\tZ}} \gamma_{{\tY\tZ}}
    + 2\gamma_{{\tX\tY\tZ}} \left(2 \sm{\gamma_\tX \gamma_\tY}+7 \sm{\gamma_{{\tX\tZ}} \gamma_\tY}\right) & \nonumber \\
   &+ 8\sm{\gamma_{{\tX\tZ}} \left(\gamma_\tX+\gamma_\tY\right) \gamma_{{\tY\tZ}}}
    + 6\sm{\gamma_{{\tX\tZ}}^2 \gamma_\tY} & \nonumber \\
   &+ 4\sm{\gamma_{{\tX\tY}} \left(\gamma_\tX+\gamma_\tY\right) \gamma_\tZ}
    +  \sm{\gamma_{{\tX\tZ}} \gamma_\tY^2} & \nonumber \\
   &+ 3\gamma_\tX \gamma_\tY \gamma_\tZ
     ,& \label{eqn:results:d32} \\
  d_{42} &=
      8\sm{\gamma_\tX \gamma_{{\tX\tZ}} \gamma_\tY \gamma_{{\tY\tZ}}}
    + 2\sm{\gamma_{{\tX\tZ}}^2 \gamma_\tY^2} & \nonumber \\
   &+  \gamma_\tX \gamma_\tY \gamma_\tZ \left(\sm{\gamma_\tX}+4 \sm{\gamma_{{\tX\tY}}}\right)
     ,& \label{eqn:results:d42} \\
  d_{44} &=
      3\gamma_\tX \gamma_\tY \gamma_\tZ \left(\sm{\gamma_\tX}+4 \sm{\gamma_{{\tX\tY}}}+8 \gamma_{{\tX\tY\tZ}}\right) & \nonumber \\
   &+22\sm{\gamma_\tX \gamma_{{\tX\tZ}} \gamma_\tY \gamma_{{\tY\tZ}}}
    + 5\sm{\gamma_{{\tX\tZ}}^2 \gamma_\tY^2} & \nonumber \\
   &+ 6\sm{\gamma_{{\tX\tY}} \left(\gamma_\tX+\gamma_\tY\right) \gamma_\tZ^2}
     ,& \label{eqn:results:d44} \\
  d_{54} &=
      3\gamma_\tX \gamma_\tY \gamma_\tZ \left(\sm{\gamma_\tX \gamma_\tY}+4 \sm{\gamma_{{\tX\tZ}} \gamma_\tY}\right)
     ,& \label{eqn:results:d54} \\
  d_{66} &=
     15\gamma_\tX^2 \gamma_\tY^2 \gamma_\tZ^2
     .& \label{eqn:results:d66}
\end{align}

\subsection{4-cumulant\label{subsec:4-cumulant}}

The four fragments of interest are $\tX$, $\tY$, $\tZ$, $\tU$. The 4-cumulant
$\kappa_4$ and its variance are given by:
\begin{align}
  \kappa_4 &= 
       \nu_0            \gamma_{\tX\tY\tZ\tU} & \nonumber \\
   &+  \nu_0^2 \sigma^2 \left(\sm{\gamma_\tU \gamma_{{\tX\tY\tZ}}}+\sm{\gamma_{{\tX\tZ}} \gamma_{{\tY\tU}}}\right)
     ,& \label{eqn:results:kappa4} \\
  \var\kappa_4
   &=  \nu_0            \gamma_{\tX\tY\tZ\tU} 
    +  \nu_0^2          e_{20}
    +  \nu_0^3          e_{30}
    +  \nu_0^4          e_{40}
    +  \nu_0^2 \sigma^2 e_{22} & \nonumber \\
   &+  \nu_0^3 \sigma^2 e_{32}
    +  \nu_0^4 \sigma^2 e_{42}
    +  \nu_0^5 \sigma^2 e_{52}
    +  \nu_0^4 \sigma^4 e_{44} & \nonumber \\
   &+  \nu_0^5 \sigma^4 e_{54} 
    +  \nu_0^6 \sigma^4 e_{64}
    +  \nu_0^6 \sigma^6 e_{66} & \nonumber \\
   &+  \nu_0^7 \sigma^6 e_{76} 
    +  \nu_0^8 \sigma^8 e_{88}
     .& \label{eqn:results:varkappa4}
\end{align}
Again, the subscripts $a$ and $b$ in $e_{ab}$ refer to the powers of $\nu_0$ and $\sigma$, respectively.
The coefficients $e_{ab}$ are:
\begin{align}
  e_{20} &=
      7\gamma_{\tX\tY\tZ\tU}^2
    +  \gamma_{\tX\tY\tZ\tU} \left(4 \sm{\gamma_{\tX\tU}}+8 \sm{\gamma_{\tX\tY\tU}}\right) & \nonumber \\
   &+ 4\sm{\gamma_{\tX\tY\tU} \gamma_{\tX\tY\tZ}}
    + 2\sm{\gamma_{\tX\tU} \gamma_{\tX\tY\tZ}}
    +  \sm{\gamma_{\tX\tZ} \gamma_{\tY\tU}} & \nonumber \\
   &+  \sm{\gamma_\tU \gamma_{\tX\tY\tZ}}
     ,& \label{eqn:results:e20} \\
  e_{30} &=
     14\gamma_{\tX\tY\tZ\tU} \sm{\gamma_{\tX\tZ} \gamma_{\tY\tU}}
    +16\sm{\gamma_{\tX\tY\tZ} \gamma_{\tX\tZ\tU} \gamma_{\tY\tU}} & \nonumber \\
   &+ 4\sm{\gamma_\tU \gamma_{\tX\tY\tZ}^2}
    + 8\sm{\gamma_{\tX\tZ\tU} \left(\gamma_{\tX\tY} \gamma_{\tY\tU}+\gamma_{\tX\tU} \gamma_{\tY\tZ}\right)} & \nonumber \\
   &+ 4\sm{\gamma_{\tX\tU} \gamma_{\tX\tZ} \gamma_{\tY\tU}}
    + 4\sm{\gamma_\tU \gamma_{\tX\tY} \gamma_{\tX\tY\tZ}} & \nonumber \\
   &+ 2\sm{\gamma_{\tX\tZ}^2 \gamma_{\tY\tU}}
    + 2\sm{\gamma_\tU \gamma_{\tX\tY} \gamma_{\tX\tZ}} & \nonumber \\
   &+  \sm{\gamma_\tU \gamma_{\tX\tZ} \gamma_\tY}
     ,& \label{eqn:results:e30} \\
  e_{40} &=
      3\sm{\gamma_{\tX\tZ}^2 \gamma_{\tY\tU}^2}
    +14\sm{\gamma_{\tX\tU} \gamma_{\tX\tZ} \gamma_{\tY\tU} \gamma_{\tY\tZ}} & \nonumber \\
   &+ 8\sm{\gamma_\tU \gamma_{\tX\tY} \gamma_{\tX\tZ} \gamma_{\tY\tZ}}
    + 2\sm{\gamma_\tU \gamma_{\tX\tZ}^2 \gamma_\tY} & \nonumber \\
   &+  \gamma_\tU \gamma_\tX \gamma_\tY \gamma_\tZ
     ,& \label{eqn:results:e40} \\
  e_{22} &=
      8\gamma_{\tX\tY\tZ\tU}^2
    + 2\gamma_{\tX\tY\tZ\tU} \left(\sm{\gamma_\tU}+2\sm{\gamma_{\tX\tU}}+4\sm{\gamma_{\tX\tY\tU}}\right) & \nonumber \\
   &+ 4\sm{\gamma_{\tX\tY\tU} \gamma_{\tX\tY\tZ}}
    + 2\sm{\gamma_{\tX\tU} \gamma_{\tX\tY\tZ}} & \nonumber \\
   &+  \sm{\gamma_{\tX\tZ} \gamma_{\tY\tU}}
    +  \sm{\gamma_\tU \gamma_{\tX\tY\tZ}}
     ,& \label{eqn:results:e22} \\
  e_{32} &=
     46\gamma_{\tX\tY\tZ\tU} \sm{\gamma_{\tX\tZ} \gamma_{\tY\tU}}
    +48\sm{\gamma_{\tX\tY\tZ} \gamma_{\tX\tZ\tU} \gamma_{\tY\tU}} & \nonumber \\
   &+ 2\gamma_{\tX\tY\tZ\tU} \left(2 \sm{\gamma_\tU \gamma_\tX}+8 \sm{\gamma_\tU \gamma_{\tX\tY}}+15 \sm{\gamma_\tU \gamma_{\tX\tY\tZ}}\right) & \nonumber \\
   &+16\sm{\gamma_\tU \gamma_{\tX\tY\tU} \gamma_{\tX\tY\tZ}}
    +12\sm{\gamma_\tU \gamma_{\tX\tY\tZ}^2} & \nonumber \\
   &+24\sm{\gamma_{\tX\tZ\tU} \left(\gamma_{\tX\tY} \gamma_{\tY\tU}+\gamma_{\tX\tU} \gamma_{\tY\tZ}\right)}
    +12\sm{\gamma_{\tX\tU} \gamma_{\tX\tZ} \gamma_{\tY\tU}} & \nonumber \\
   &+ 6\sm{\gamma_{\tX\tZ}^2 \gamma_{\tY\tU}}
    + 8\sm{\gamma_\tU \gamma_{\tX\tU} \gamma_{\tX\tY\tZ}}
    +12\sm{\gamma_\tU \gamma_{\tX\tY} \gamma_{\tX\tY\tZ}} & \nonumber \\
   &+ 8\sm{\gamma_\tU \gamma_{\tX\tY\tU} \gamma_{\tX\tZ}}
    + 6\sm{\gamma_\tU \gamma_{\tX\tY} \gamma_{\tX\tZ}} 
    +  \sm{\gamma_\tU^2 \gamma_{\tX\tY\tZ}} & \nonumber \\
   &+ 4\sm{\gamma_\tU \gamma_{\tX\tZ} \gamma_{\tY\tU}}
    + 4\sm{\gamma_\tU \gamma_\tX \gamma_{\tX\tY\tZ}} & \nonumber \\
   &+ 3\sm{\gamma_\tU \gamma_{\tX\tZ} \gamma_\tY}
     ,& \label{eqn:results:e32} \\
  e_{42} &=
     46\sm{\gamma_\tU \gamma_{\tX\tY\tZ}} \sm{\gamma_{\tX\tZ} \gamma_{\tY\tU}}
    +16\sm{\gamma_\tU \gamma_{\tX\tY\tZ} \gamma_{\tX\tZ\tU} \gamma_\tY} & \nonumber \\
   &+48\sm{\gamma_\tU \gamma_{\tX\tY\tU} \gamma_{\tX\tZ} \gamma_{\tY\tZ}} 
    +92\sm{\gamma_{\tX\tU} \gamma_{\tX\tZ} \gamma_{\tY\tU} \gamma_{\tY\tZ}} & \nonumber \\
   &+14\gamma_{\tX\tY\tZ\tU} \sm{\gamma_\tU \gamma_{\tX\tZ} \gamma_\tY}
    + 4\sm{\gamma_\tU^2 \gamma_{\tX\tY\tZ}^2} & \nonumber \\
   &+22\sm{\gamma_{\tX\tZ}^2 \gamma_{\tY\tU}^2}
    + 8\sm{\gamma_\tU \gamma_{\tX\tU} \gamma_{\tX\tY\tZ} \gamma_\tY} & \nonumber \\
   &+ 4\sm{\gamma_\tU^2 \gamma_{\tX\tY} \gamma_{\tX\tY\tZ}}
    +24\sm{\gamma_\tU \gamma_{\tX\tY} \gamma_{\tX\tZ} \gamma_{\tY\tU}} & \nonumber \\
   &+48\sm{\gamma_\tU \gamma_{\tX\tY} \gamma_{\tX\tZ} \gamma_{\tY\tZ}}
    + 8\sm{\gamma_\tU \gamma_{\tX\tY\tU} \gamma_{\tX\tZ} \gamma_\tY} & \nonumber \\
   &+24\sm{\gamma_\tU \gamma_{\tX\tY\tZ} \gamma_{\tX\tZ} \gamma_\tY}
    +12\sm{\gamma_\tU \gamma_{\tX\tZ}^2 \gamma_{\tY\tU}} & \nonumber \\
   &+ 4\sm{\gamma_\tU \gamma_\tX \gamma_{\tX\tY\tZ} \gamma_\tY}
    + 4\sm{\gamma_\tU \gamma_\tX \gamma_{\tX\tZ} \gamma_{\tY\tU}} & \nonumber \\
   &+12\sm{\gamma_\tU \gamma_{\tX\tU} \gamma_{\tX\tZ} \gamma_\tY}
    + 2\sm{\gamma_\tU^2 \gamma_{\tX\tY} \gamma_{\tX\tZ}} & \nonumber \\
   &+ 4\sm{\gamma_\tU \gamma_{\tX\tZ} \gamma_\tY \gamma_{\tY\tU}}
    +12\sm{\gamma_\tU \gamma_{\tX\tZ}^2 \gamma_\tY} & \nonumber \\
   &+ 6\sm{\gamma_\tU \gamma_\tX \gamma_{\tX\tZ} \gamma_\tY}
    +  \sm{\gamma_\tU^2 \gamma_{\tX\tZ} \gamma_\tY} & \nonumber \\
   &+ 6\gamma_\tU \gamma_\tX \gamma_\tY \gamma_\tZ
     ,& \label{eqn:results:e42} \\
  e_{52} &=
     14\sm{\gamma_\tU \gamma_{\tX\tU} \gamma_{\tX\tZ} \gamma_\tY \gamma_{\tY\tZ}}
    + 8\sm{\gamma_\tU^2 \gamma_{\tX\tY} \gamma_{\tX\tZ} \gamma_{\tY\tZ}} & \nonumber \\
   &+ 6\sm{\gamma_\tU \gamma_{\tX\tZ}^2 \gamma_\tY \gamma_{\tY\tU}}
    + 8\sm{\gamma_\tU \gamma_\tX \gamma_{\tX\tZ} \gamma_\tY \gamma_{\tY\tZ}} & \nonumber \\
   &+ 2\sm{\gamma_\tU^2 \gamma_{\tX\tZ}^2 \gamma_\tY}
    + 4\sm{\gamma_\tU \gamma_\tX \gamma_{\tX\tU} \gamma_\tY \gamma_\tZ}  & \nonumber \\
   &+  \gamma_\tU \gamma_\tX \gamma_\tY \gamma_\tZ \sm{\gamma_\tU}
     ,& \label{eqn:results:e52} \\
  e_{44} &=
     24\gamma_{\tX\tY\tZ\tU} \left(\sm{\gamma_\tU \gamma_\tX \gamma_\tY}+2 \sm{\gamma_\tU \gamma_{\tX\tZ} \gamma_\tY}\right) & \nonumber \\
   &+48\sm{\gamma_\tU \gamma_{\tX\tY\tU} \gamma_{\tX\tZ} \gamma_{\tY\tZ}}
    +46\sm{\gamma_\tU \gamma_{\tX\tY\tZ}} \sm{\gamma_{\tX\tZ} \gamma_{\tY\tU}} & \nonumber \\
   &+46\sm{\gamma_\tU \gamma_{\tX\tY\tZ} \gamma_{\tX\tZ\tU} \gamma_\tY}
    +46\sm{\gamma_{\tX\tU} \gamma_{\tX\tZ} \gamma_{\tY\tU} \gamma_{\tY\tZ}} & \nonumber \\
   &+11\sm{\gamma_\tU^2 \gamma_{\tX\tY\tZ}^2}
    +11\sm{\gamma_{\tX\tZ}^2 \gamma_{\tY\tU}^2}
    +12\sm{\gamma_\tU^2 \gamma_{\tX\tY} \gamma_{\tX\tY\tZ}} & \nonumber \\
   &+24\sm{\gamma_\tU \gamma_{\tX\tU} \gamma_{\tX\tY\tZ} \gamma_\tY}
    +24\sm{\gamma_\tU \gamma_{\tX\tY} \gamma_{\tX\tZ} \gamma_{\tY\tU}} & \nonumber \\
   &+24\sm{\gamma_\tU \gamma_{\tX\tY} \gamma_{\tX\tZ} \gamma_{\tY\tZ}}
    +24\sm{\gamma_\tU \gamma_{\tX\tY\tU} \gamma_{\tX\tZ} \gamma_\tY} & \nonumber \\
   &+24\sm{\gamma_\tU \gamma_{\tX\tY\tZ} \gamma_{\tX\tZ} \gamma_\tY}
    +12\sm{\gamma_\tU \gamma_{\tX\tZ}^2 \gamma_{\tY\tU}} & \nonumber \\
   &+12\sm{\gamma_\tU \gamma_\tX \gamma_{\tX\tY\tZ} \gamma_\tY}
    +12\sm{\gamma_\tU \gamma_\tX \gamma_{\tX\tZ} \gamma_{\tY\tU}} & \nonumber \\
   &+ 6\sm{\gamma_\tU^2 \gamma_\tX \gamma_{\tX\tY\tZ}}
    +12\sm{\gamma_\tU \gamma_{\tX\tU} \gamma_{\tX\tZ} \gamma_\tY} & \nonumber \\
   &+ 6\sm{\gamma_\tU^2 \gamma_{\tX\tY} \gamma_{\tX\tZ}}
    +12\sm{\gamma_\tU \gamma_{\tX\tZ} \gamma_\tY \gamma_{\tY\tU}} & \nonumber \\
   &+ 6\sm{\gamma_\tU \gamma_{\tX\tZ}^2 \gamma_\tY}
    + 6\sm{\gamma_\tU \gamma_\tX \gamma_{\tX\tZ} \gamma_\tY} & \nonumber \\
   &+ 3\sm{\gamma_\tU^2 \gamma_{\tX\tZ} \gamma_\tY}
    + 3\gamma_\tU \gamma_\tX \gamma_\tY \gamma_\tZ
     ,& \label{eqn:results:e44} \\
  e_{54} &=
       \gamma_\tU \gamma_\tX \gamma_\tY \gamma_\tZ \left(9 \sm{\gamma_\tU}+48 \sm{\gamma_{\tX\tY\tU}}+42 \gamma_{\tX\tY\tZ\tU}\right) & \nonumber \\
   &+94\sm{\gamma_\tU \gamma_\tX \gamma_{\tX\tZ\tU} \gamma_\tY \gamma_{\tY\tZ}}
   +142\sm{\gamma_\tU \gamma_{\tX\tU} \gamma_{\tX\tZ} \gamma_\tY \gamma_{\tY\tZ}} & \nonumber \\
   &+72\sm{\gamma_\tU^2 \gamma_{\tX\tY} \gamma_{\tX\tZ} \gamma_{\tY\tZ}}
    +46\sm{\gamma_\tU^2 \gamma_{\tX\tY\tZ} \gamma_{\tX\tZ} \gamma_\tY} & \nonumber \\
   &+70\sm{\gamma_\tU \gamma_{\tX\tZ}^2 \gamma_\tY \gamma_{\tY\tU}}
    +12\sm{\gamma_\tU^2 \gamma_\tX \gamma_{\tX\tY\tZ} \gamma_\tY} & \nonumber \\
   &+48\sm{\gamma_{\tX\tZ} \gamma_{\tY\tU}} \sm{\gamma_\tU \gamma_\tX \gamma_\tY}
    +72\sm{\gamma_\tU \gamma_\tX \gamma_{\tX\tZ} \gamma_\tY \gamma_{\tY\tZ}} & \nonumber \\
   &+24\sm{\gamma_\tU^2 \gamma_{\tX\tY} \gamma_{\tX\tZ} \gamma_\tY}
    +18\sm{\gamma_\tU^2 \gamma_{\tX\tZ}^2 \gamma_\tY} & \nonumber \\
   &+ 3\sm{\gamma_\tU^2 \gamma_{\tX\tZ} \gamma_\tY^2}
    +36\sm{\gamma_\tU \gamma_\tX \gamma_{\tX\tU} \gamma_\tY \gamma_\tZ} & \nonumber \\
   &+12\sm{\gamma_\tU^2 \gamma_\tX \gamma_{\tX\tZ} \gamma_\tY}
     ,& \label{eqn:results:e54} \\
  e_{64} &=
     40\gamma_\tU \gamma_\tX \gamma_\tY \gamma_\tZ \sm{\gamma_{\tX\tZ} \gamma_{\tY\tU}}
    +22\sm{\gamma_\tU^2 \gamma_\tX \gamma_{\tX\tZ} \gamma_\tY \gamma_{\tY\tZ}} & \nonumber \\
   &+ 5\sm{\gamma_\tU^2 \gamma_{\tX\tZ}^2 \gamma_\tY^2} & \nonumber \\
   &+  \gamma_\tU \gamma_\tX \gamma_\tY \gamma_\tZ \left(3 \sm{\gamma_\tU \gamma_\tX}+12 \sm{\gamma_\tU \gamma_{\tX\tY}}\right)
     ,& \label{eqn:results:e64} \\
  e_{66} &=
      6\gamma_\tU \gamma_\tX \gamma_\tY \gamma_\tZ \left(19\sm{\gamma_\tU \gamma_{\tX\tY\tZ}}+39\sm{\gamma_{\tX\tZ} \gamma_{\tY\tU}}\right) & \nonumber \\
  &+120\sm{\gamma_\tU^2 \gamma_\tX \gamma_{\tX\tZ} \gamma_\tY \gamma_{\tY\tZ}}
    +30\sm{\gamma_\tU^2 \gamma_{\tX\tZ}^2 \gamma_\tY^2} & \nonumber \\
  &+ 15\gamma_\tU \gamma_\tX \gamma_\tY \gamma_\tZ \left(\sm{\gamma_\tU \gamma_\tX}+4\sm{\gamma_\tU \gamma_{\tX\tY}} \right) & \nonumber \\
  & +30\sm{\gamma_\tU^2 \gamma_\tX \gamma_{\tX\tZ} \gamma_\tY^2}
     ,& \label{eqn:results:e66} \\
  e_{76} &=
     54\gamma_\tU \gamma_\tX \gamma_\tY \gamma_\tZ \sm{\gamma_\tU \gamma_{\tX\tZ} \gamma_\tY} & \nonumber \\
  &+ 15\gamma_\tU \gamma_\tX \gamma_\tY \gamma_\tZ \sm{\gamma_\tU \gamma_\tX \gamma_\tY}
     ,& \label{eqn:results:e76} \\
  e_{88} &=
     96\gamma_\tU^2 \gamma_\tX^2 \gamma_\tY^2 \gamma_\tZ^2
     .& \label{eqn:results:e88}
\end{align}

\section{Discussion\label{sec:discussion}}

The two key consequences of the imperfection of the measurement apparatus on a
cumulant-mapping measurement are immediately apparent from the general results
above: First, the noise-induced contribution to the cumulant $\kappa_n$ always
appears in the second order of the average event rate $\nu_0$. For a given
level of noise, there always exists a critical event rate $\nu_{n,\text{crit}}$,
which should not be exceeded, to avoid contamination of the results:
\begin{align}
  \nu_{2,\text{crit}} &= \frac{\epsilon}{\sigma^2} \frac{\gamma_{\tX\tY}}
                   {\gamma_\tX\gamma_\tY} & \label{eqn:discussion:nu2c} \\
  \nu_{3,\text{crit}} &= \frac{\epsilon}{\sigma^2} \frac{\gamma_{\tX\tY\tZ}}
                   {\sm{\gamma_{\tX\tZ}\gamma_\tY}} & \label{eqn:discussion:nu3c} \\
  \nu_{4,\text{crit}} &= \frac{\epsilon}{\sigma^2} \frac{\gamma_{\tX\tY\tZ\tU}}
                   {\sm{\gamma_\tU \gamma_{{\tX\tY\tZ}}}+\sm{\gamma_{{\tX\tZ}} \gamma_{{\tY\tU}}}} & \label{eqn:discussion:nu4c} 
\end{align}
where $\epsilon$ is the permissible fraction of the false-cumulant events
relative to the true-cumulant signal. 

Second, the external noise could potentially lead to an increase in the width
of the distribution, given by the square root of the corresponding variance,
from which the cumulant is sampled. In a perfect apparatus, the width of the
distribution for $\kappa_n$ grows generally as $\nu_0^{n/2}$ with the average
event rate.  If no restrictions are imposed on the event rate, the
fastest-growing noise-induced contribution raises as $\nu_0^n\sigma^n$, and
will eventually dominate the distribution width.  If the event rate is held
under the critical $\nu_{n,\text{crit}}$, the product $\nu_0\sigma^2$ is
bounded by a constant (See
\crefrange{eqn:discussion:nu2c}{eqn:discussion:nu4c}), and the effective
noise-induced width grows as $\nu_0^{n/2}$ as well. The final distribution
width then depends on the specific parameters.

The high-order dependence of the distribution width for $\kappa_n$ with $n\ge3$
on the event rate leads to a paradoxical situation in a cumulant-mapping
measurement, where higher event rate increases the number of events observed in
a given length of time -- yet the measurement becomes \textit{less} accurate,
due to the faster broadening of the cumulant distribution.

We will now turn to some specific, representative measurement regimes. We will
consider three scenarios: A dominant process with no background
(subsection~\ref{sec:discussion:dominant}), a minor process on a strong
background (subsection~\ref{sec:discussion:minor}), and a minor process, where
\textit{one} of the fragments is background-free
(subsection~\ref{sec:discussion:marker}). For simplicity, and unless stated
otherwise, we will assume unit detection efficiency throughout
($\beta_\tX=\beta_\tY=\beta_\tZ=\beta_\tU=1$). Due to the linear relationship
between the intrinsic ($\alpha$) and detector-based ($p$) elementary-event 
probabilities, a choice of $\beta<1$ amounts to a redefinition of the 
events, provided that none of the detection probabilities vanish -- see the 
discussion following Eq.~\eqref{eqn:notation:p} -- and leaves the conclusions
qualitatively unaffected. Furthermore, we set both the permissible
contamination threshold $\epsilon$ and the noise level $\sigma$ at $1\%$,
again unless stated otherwise.

\subsection{A dominant process\label{sec:discussion:dominant}}

\begin{figure}[!tbhp]
  \includegraphics[trim=0pt 0pt 0pt 0pt,clip,width=\figurewidth]{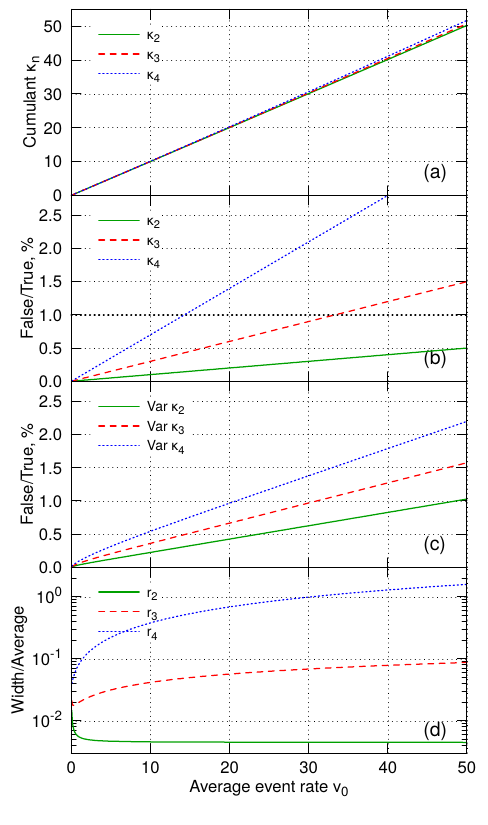}
  \caption{\label{fig:case-a} 
Dominant process with no background (section~\ref{sec:discussion:dominant}),
$\gamma_a=1$, noise level $\sigma=1\%$, $N=10^5$ shots. Green solid line:
covariance (2-cumulant) $\kappa_2$. Red dashed line: 3-cumulant $\kappa_3$.
Blue dotted line: 4-cumulant $\kappa_4$. Panel (a): total value of the
cumulant, as a function of the average event rate $\nu_0$. Panel (b): the ratio
of the false (noise-induced) and true contributions to the cumulant in percent.
Panel (c): the ratio of the false and true contributions to the cumulant
variance. Panel (d): Measurement uncertainty $r_n$, \cref{eqn:discussion:ratio}.
}
\end{figure}

Here, we assume that every primitive event produces all fragments of interest,
with a unit probability. All these fragments are reliably detected, and no
other fragmentation processes occur in the system. This scenario could be seen
as an idealized model of the Coulomb explosion. Under our assumptions, all
$\gamma$ parameters take unit value (See \cref{eqn:notation:gamma}).
\Cref{eqn:results:kappa2,eqn:results:varkappa2,eqn:results:kappa3,eqn:results:varkappa3,eqn:results:kappa4,eqn:results:varkappa4}
then take the much-simplified form:
\begin{align}
  \kappa_2 &\overset{\text{dm}}{=}
     \nu_0 + \nu_0^2 \sigma^2
     ,& \label{eqn:discussion:dominant:kappa2} \\
  \kappa_3 &\overset{\text{dm}}{=}
     \nu_0 + 3 \nu_0^2 \sigma^2
     ,& \label{eqn:discussion:dominant:kappa3} \\
  \kappa_4 &\overset{\text{dm}}{=}
     \nu_0 + 7 \nu_0^2 \sigma^2
     ,& \label{eqn:discussion:dominant:kappa4} \\
  \var\kappa_2 &\overset{\text{dm}}{=}
     \nu_0 +2 \nu_0^2 +7 \nu_0^2 \sigma^2 +4 \nu_0^3 \sigma^2 +2 \nu_0^4 \sigma^4
     ,& \label{eqn:discussion:dominant:varkappa2} \\
  \var\kappa_3 &\overset{\text{dm}}{=}
      \nu_0 +24 \nu_0^2 +15 \nu_0^3 +31 \nu_0^2 \sigma^2 & \nonumber \\
    & +174 \nu_0^3 \sigma^2 +45 \nu_0^4 \sigma^2 +186 \nu_0^4 \sigma^4 & \nonumber \\
    & +45 \nu_0^5 \sigma^4 +15 \nu_0^6 \sigma^6
     ,& \label{eqn:discussion:dominant:varkappa3} \\
  \var\kappa_4 &\overset{\text{dm}}{=}
      \nu_0 +118 \nu_0^2 +484 \nu_0^3 +96 \nu_0^4 +127 \nu_0^2 \sigma^2 & \nonumber \\
    & +2380 \nu_0^3 \sigma^2 +3796 \nu_0^4 \sigma^2 +5054 \nu_0^4 \sigma^4 & \nonumber \\
    & +384 \nu_0^5 \sigma^2 +7260 \nu_0^5 \sigma^4 +576 \nu_0^6 \sigma^4 & \nonumber \\
    & +3948 \nu_0^6 \sigma^6 +384 \nu_0^7 \sigma^6 +96 \nu_0^8 \sigma^8
     .& \label{eqn:discussion:dominant:varkappa4} 
\end{align}
(The overscript ``dm'' for ``dominant'' serves to emphasize that 
\crefrange{eqn:discussion:dominant:kappa2}{eqn:discussion:dominant:varkappa4}
only apply in this special case.)

The key properties of the covariant mapping in this case are illustrated in
Figure~\ref{fig:case-a}, for $\kappa_n, n=2,3,4$. In all three cases, the
expectation of the cumulant grows almost perfectly linearly with the average
event rate, at least until $\nu_0=50$ (Panel a). The contribution from the
noise-induced, false covariance remains small, on the level of a few percent
(Panel b). Our chosen contamination tolerance of 1\% is reached at
$\nu_{4,\text{crit}}=14.3$, $\nu_{3,\text{crit}}=33.3$, and
$\nu_{2,\text{crit}}=100$. If higher contamination levels are permissible, then
even higher even rates are possible. The noise contribution to the variance
is even smaller at these events rates (Panel c). 

A practically important property of the cumulant is the ratio of the 
width of the distribution of the sampled expectation and the expectation
itself, which determines the experimental accuracy. For a measurement with
$N$ shots, the relative width $r_{n}$ is given by:
\begin{align}
  r_{n} &= \frac{1}{\kappa_n}\sqrt{\frac{1}{N}\var\kappa_n}. & \label{eqn:discussion:ratio}
\end{align}
This quantity is plotted in Figure~\ref{fig:case-a}(d) for $N=10^5$. As expected,
the $2-$cumulant (the covariance) shows a behavior qualitatively different from
the higher cumulants: The $r_2$ decreases, essentially monotonically, reaching
the asymptote of about $0.014$ at $\nu_0>50$. In contrast, the $r_3$ and $r_4$
reach a minimum at, respectively, $\nu_0=0.26$ and $\nu_0=0.045$.
Asymptotically, both grow without a bound, respectively linearly and
quadratically with $\nu_0$. At the critical event rate $\nu_{n,\text{crit}}$,
we obtain $r_3=0.23$ and $r_4=1.6$, so that the width of the distribution is
comparable to the magnitude of the cumulant.  As the result, a number of shots
much higher than $10^5$ is likely necessary to obtain accurate results for
higher cumulants in this case (See Section~\ref{sec:convergence} below).

Overall, the Coulomb explosion appears to be the ideal case for applications of
the cumulant mapping\cite{cheng2023:prl}, even in the presence of moderate noise levels,
provided that it remains the dominant fragmentation channel.

\subsection{A minor process\label{sec:discussion:minor}}

\begin{figure}[!tbhp]
  \includegraphics[trim=0pt 0pt 0pt 0pt,clip,width=\figurewidth]{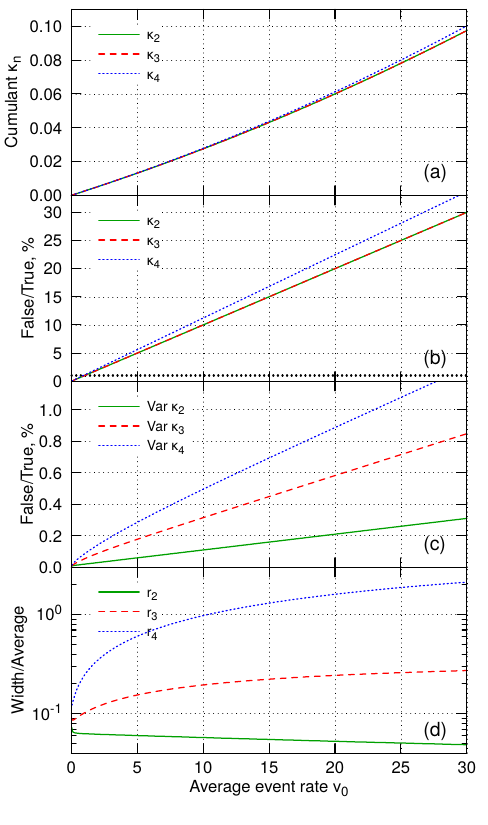}
  \caption{\label{fig:case-b} 
Minor process with strong, correlated background
(section~\ref{sec:discussion:minor}), $\tau=2.5\times10^{-3}$, noise level
$\sigma=1\%$, $N=10^7$ shots. Also see Fig.~\ref{fig:case-a} for panel
description.
}
\end{figure}

Our second hypothetical scenario involves a minor channel of interest, which
occurs in a fraction $\tau$ ($\tau\ll1$) of the primitive fragmentation events.
For illustrative purposes, we choose $\tau=0.25\%$.  All other fragmentation
channels are assumed to occur with equal probability. For example, for the
3-cumulant, primitive fragmentation channels leading to fragments $\tX$, $\tY$,
$\tZ$, $\tX+\tY$, $\tX+\tZ$, and $\tY+\tZ$ are taken to be equally probable, at
$\alpha=\frac{1}{6}\left(1-\tau\right)\approx16.6\%$ The number of shots is now
$N=10^7$. For our choice, the background is \textit{partially} correlated:
The events producing $\tX$, $\tY$, and $\tZ$ alone form the uncorrelated part of 
the background, as considered by Frasinski\cite{Frasinski22a}. On the other hand,
events producing fragment pairs form the correlated background. Both correlated
and uncorrelated background induce false-cumulant contributions in the presence
of noise, but, as will be seen shortly, with a dramatically different efficiency.

Keeping only the leading non-zero term in the first-order $\tau$ expansion of each 
$\nu_0^a\sigma^b$ prefactor\cite{supplementary}, the cumulants are now given by:
\begin{align}
  \kappa_2 &\overset{\text{mc}}{\approx}
     \nu_0 \tau + \frac{1}{4} \nu_0^2 \sigma^2
     ,& \label{eqn:discussion:minor-c:kappa2} \\
  \kappa_3 &\overset{\text{mc}}{\approx}
     \nu_0 \tau + \frac{1}{4} \nu_0^2 \sigma^2
     ,& \label{eqn:discussion:minor-c:kappa3} \\
  \kappa_4 &\overset{\text{mc}}{\approx}
     \nu_0 \tau + \frac{55}{196} \nu_0^2 \sigma^2
     .& \label{eqn:discussion:minor-c:kappa4}
\end{align}
(The overscript ``mc'' stands for ``minor, correlated'', emphasizing the
special-case nature of the equations.)

The true-cumulant contribution, linear in $\nu_0$, is proportional to the
primitive event rate $\tau$, as expected. However, the false-cumulant term is
$\tau$-independent. As the result, the total cumulant [Fig.~\ref{fig:case-b}(a)]
now visibly deviates from linear dependence on $\nu_0$. The false-cumulant
contribution reaches $30\%$ already at the $\nu_0=30$ average event rate. Our
chosen critical threshold of $1\%$ false-cumulant contamination is reached at
$\nu_0\approx1$, in all three cases. At such low event rates, coincidence
detection, which is immune to noise for high detection
efficiencies\cite{Mikosch13a}, is likely the preferred detection mode.

The full expressions for $\var{\kappa_n}$ are somewhat
lengthy\cite{supplementary}, but for our choice of parameters and $\nu_0\le30$,
they are adequately approximated by:
\begin{align}
  \var\kappa_2 &\overset{\text{mc}}{\approx}
      0.250 \nu_0^2
     ,& \label{eqn:discussion:minor-c:varkappa2} \\
  \var\kappa_3 &\overset{\text{mc}}{\approx}
      0.417 \nu_0^2 + 0.246 \nu_0^3
     ,& \label{eqn:discussion:minor-c:varkappa3} \\
  \var\kappa_4 &\overset{\text{mc}}{\approx}
      0.770 \nu_0^2 +2.61 \nu_0^3 +0.467 \nu_0^4
     .& \label{eqn:discussion:minor-c:varkappa4} 
\end{align}
As before, the variances are dominated by the true-cumulant contribution, with
less than $1.5\%$ stemming from the noise for $\nu_0\le30$
[Fig.~\ref{fig:case-b}(c)]. Finally, the relative width of the distribution
[Fig.~\ref{fig:case-b}(d)] follows the same trend as before: the $r_2$
monotonically decreases in the region of interest, while $r_3$
($r_4$) reaches a minimum at $\nu_0=0.10$ ($\nu_0=0.12$) and
increases linearly (quadratically) for large $\nu_0$. 

This scenario is clearly unfavorable for cumulant detection, with either very
low average event rates, or exceptionally high stability of the experimental
setup being essential. The reason behind this somewhat disappointing outcome
is the partially-correlated nature of the dominant background. From
\cref{eqn:results:kappa3}, the noise combines the one- and two-fragment
correlations into false 3-particle cumulant. Similarly, the one- and three- and
two separate two-particle correlations are noise-coupled to produce false
4-cumulant (\cref{eqn:results:kappa4}). When the background is already
partially-correlated, even small levels of noise are sufficient to swamp the
weak signal of interest.

\begin{figure}[!tbhp]
  \includegraphics[trim=0pt 0pt 0pt 0pt,clip,width=\figurewidth]{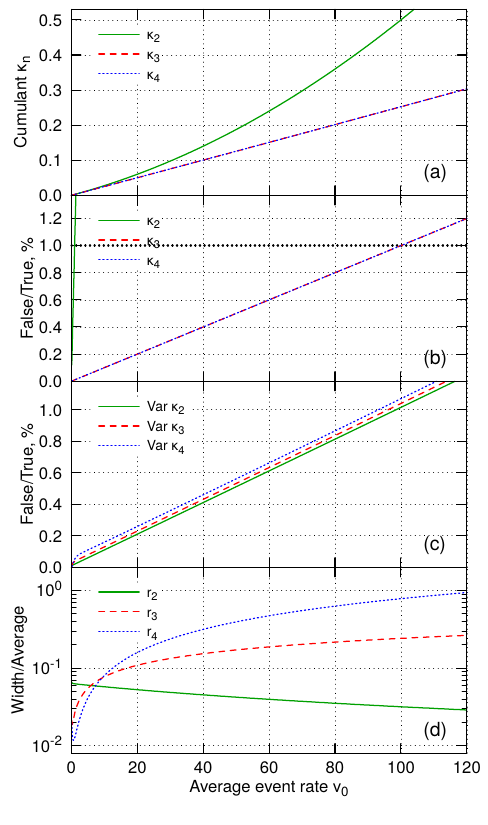}
  \caption{\label{fig:case-c} 
Minor process with strong, but uncorrelated background
(section~\ref{sec:discussion:minor}), $\tau=2.5\times10^{-3}$, noise level
$\sigma=1\%$, $N=10^7$ shots. Also see Fig.~\ref{fig:case-a} for panel
description.
}
\end{figure}

For higher cumulants (but not for the 2-particle covariance, for which the
scenario below is identical to the correlated-background case), the situation changes
dramatically if the background is strong, but uncorrelated (See
Fig.~\ref{fig:case-c}). Now, the primitive fragmentation event is assumed to
lead to either the $n$-particle fragmentation (probability $\alpha=\tau$), or to
only \textit{one} of the fragments, each with an equal probability
$\alpha=(1-\tau)/n$ ($n=2,3,4$). All other primitive-event probabilities vanish.
The 3- and 4-cumulants are now given by:
\begin{align}
  \kappa_{3,4} &\overset{\text{mu}}{\approx}
     \nu_0 \tau + \nu_0^2 \sigma^2 \tau
     ,& \label{eqn:discussion:minor-u:kappa34}
\end{align}
so that \textit{both} the true- and the false-cumulants are now proportional to
the desired-even rate $\tau$. In the relevant range of $\nu_0$, the variances
behave approximately like:
\begin{align}
  \var\kappa_3 &\overset{\text{mu}}{\approx}
     0.037 \nu_0^3
     ,& \label{eqn:discussion:minor-u:varkappa3} \\
  \var\kappa_4 &\overset{\text{mu}}{\approx}
     0.00391 \nu _0^4
     .& \label{eqn:discussion:minor-u:varkappa4} 
\end{align}
(The overscript ``mu'' stands for ``minor, uncorrelated''.)

The 3- and 4-cumulants are now essentially linear in $\nu_0$, for event rates
up to several hundred. In contrast, the $\kappa_2$ is nearly-perfectly
quadratic in this range of $\nu_0$, and is dominated by the false covariance [Fig.~\ref{fig:case-c}(a)].
The critical event rate for $\kappa_{3,4}$ is now reached at $\nu_0=100$
[Fig.~\ref{fig:case-c}(b)], while the noise-induced contributions to the
variances remain small in this range [Fig.~\ref{fig:case-c}(c)]. The behavior of
the relative width of the distribution of $\kappa$ [Fig.~\ref{fig:case-c}(d)]
remains qualitatively the same, with the optimal width reached for very low
event rates: $\nu_0=0.26$ ($\nu_0=0.65$) for $\kappa_3$ ($\kappa_4$).

Thus, the nature of the background events plays a critical role for
applications of higher-cumulant mapping to minor channels. As long as the
background is perfectly uncorrelated, the cumulants remain uncontaminated
for very high event rates. On the other hand, presence of already moderate
two-particle correlated background interferes with both 3- and
4-cumulant.

\subsection{A minor process with a marker fragment\label{sec:discussion:marker}}

\begin{figure}[!tbhp]
  \includegraphics[trim=0pt 0pt 0pt 0pt,clip,width=\figurewidth]{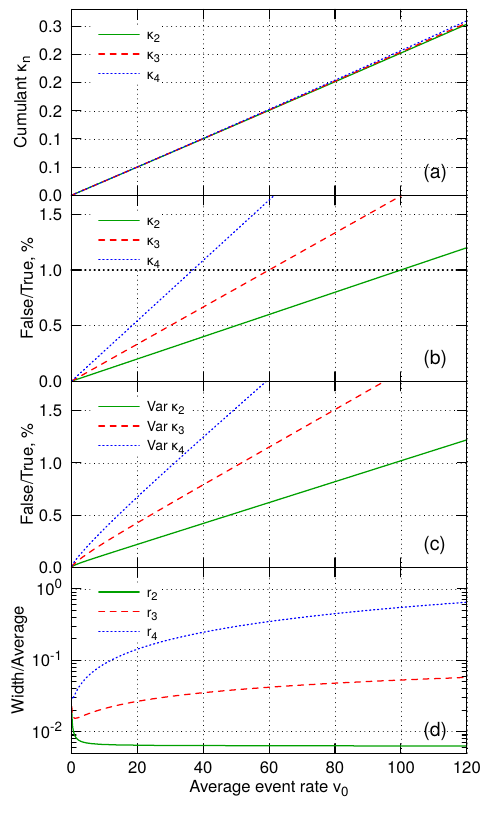}
  \caption{\label{fig:case-d} 
Minor process with strong, correlated background and one, background-free
(marker) fragment channel (section~\ref{sec:discussion:marker}),
$\tau=2.5\times10^{-3}$, noise level $\sigma=1\%$, $N=10^7$ shots. Also see
Fig.~\ref{fig:case-a} for panel description.
}
\end{figure}

Our final scenario reflects a not uncommon situation, where \textit{one} of the
fragments comes exclusively from the process of interest, and is
background-free. Other fragments appear on top of a strong, correlated
background. All present background fragmentation pathways taken as equally
likely, similar to Section~\ref{sec:discussion:minor} above. Thus, the
background-free fragment serves as a ``marker'' of the process of interest.
Under the same assumptions as in Section~\ref{sec:discussion:minor}, the
$n-$cumulants now become:
\begin{align}
  \kappa_2 &\overset{\text{mt}}{\approx}
     \nu_0 \tau + \nu_0^2 \sigma^2 \tau 
     ,& \label{eqn:discussion:marker:kappa2} \\
  \kappa_3 &\overset{\text{mt}}{\approx}
     \nu_0 \tau + \frac{5}{3} \nu_0^2 \sigma^2 \tau 
     ,& \label{eqn:discussion:marker:kappa3} \\
  \kappa_4 &\overset{\text{mt}}{\approx}
     \nu_0 \tau + \frac{19}{7} \nu_0^2 \sigma^2 \tau 
     .& \label{eqn:discussion:marker:kappa4}
\end{align}
so that both the true- and false-cumulant contributions are proportional to the
desired event's probability $\tau$. (The overscript ``mt'' stands for ``minor, tagged.'')
This is an extremely favorable situation:
the critical event rate $\nu_\text{crit}$ does not depend on how small $\tau$
is. For our chosen $1\%$ noise level and false-cumulant tolerance,
$\nu_{2,\text{crit}}=100$, $\nu_{3,\text{crit}}=60$, and $\nu_{4,\text{crit}}=37$
[See Fig.~\ref{fig:case-d}(b)].

The variances of the cumulant are also proportional to $\tau$ in this scenario.
Keeping only the terms relevant below the critical event rate, we obtain:
\begin{align}
  \var\kappa_2 &\overset{\text{mt}}{\approx}
     \nu_0^2 \tau
     ,& \label{eqn:discussion:marker:varkappa2} \\
  \var\kappa_3 &\overset{\text{mt}}{\approx}
      4.33 \nu_0^2 \tau + 0.668 \nu_0^3 \tau 
     ,& \label{eqn:discussion:marker:varkappa3} \\
  \var\kappa_4 &\overset{\text{mt}}{\approx}
     11.8 \nu_0^3 \tau + 0.658 \nu_0^4 \tau 
     .& \label{eqn:discussion:marker:varkappa4}
\end{align}
The false-cumulant contribution to the variance, again, remains small
[Fig.~\ref{fig:case-d}(c)].  For our chosen parameters, the relative widths of
the $\kappa_n$ distributions follow the same pattern seen for the
dominant-contribution case (Section~\ref{sec:discussion:dominant} above).
Quantity $\sqrt{\var\kappa_2}/\kappa_2$ goes to zero as $\nu_0^{-1/2}$ with
increasing $\nu_0$.  In contrast, $\sqrt{\var\kappa_{3,4}}/\kappa_{3,4}$ grow as
$\nu_0^{1/2}$ and $\nu_0^{3/2}$ asymptotically [Fig.~\ref{fig:case-d}(d)].  Their
minima are again found at low $\nu_0$ ($1.2$ and $0.3$, respectively).

Thus, the presence of a background-free fragment in a minor fragmentation
channel makes cumulant mapping robust with respect to moderate noise levels.

\subsection{Convergence of the cumulant maps\label{sec:convergence}}

\begin{figure}[!tbhp]
  \includegraphics[trim=0pt 0pt 0pt 0pt,clip,width=\figurewidth]{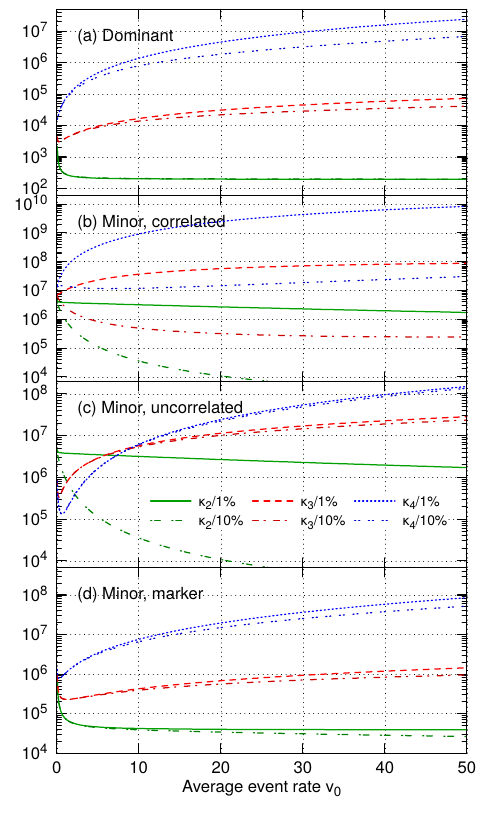}
  \caption{\label{fig:convergence} 
Number of shots required to reach $20\%$ relative convergence $\omega$ of the 
cumulant maps ($95\%$ confidence level $p$).
(a) Dominant process, no background, Fig.~\ref{fig:case-a}.
(b) Minor process, correlated background, Fig.~\ref{fig:case-b}. 
(c) Minor process, uncorrelated background, Fig.~\ref{fig:case-c}. 
(d) Minor process with a marker, Fig.~\ref{fig:case-d}.
Green solid (dot-dashed) line: 2-cumulant, $\sigma=1\%$ ($10\%$).
Red dashed (dash-dotted) line: 3-cumulant, $\sigma=1\%$ ($10\%$).
Blue dotted (double-dotted) line: 4-cumulant, $\sigma=1\%$ ($10\%$).
}
\end{figure}

A key question in planning an experimental campaign is the expected number of
shots needed to converge the observables to the desired accuracy.  The detailed
convergence dynamics depends on the specific experimental and acquisition
setup, as well as on the product channel distribution and their discrimination
on the detector.  Numerical simulations of the measurement process may be
needed to fully characterize it.  Nonetheless, our analysis above also offers
useful insight, at a much lower effort.  

Because the $n-$cumulants are sampled from a distribution with the mean
$\kappa_n$ and variance $\var{\kappa_n}$, for a sufficiently large number of
measurements $N$ the central-limit theorem\cite{DeGroot02} guarantees that the
corresponding sample average $\bar{x}_N$ is derived from a normal distribution
with the same mean and variance $N^{-1} \var{\kappa_n}$, cf.
Eq.~\eqref{eqn:discussion:ratio}. A natural way to define the convergence of
the average is to require, that the probability $P$ of $\bar{x}_N$ lying within
$\omega \kappa_n$ ($0<\omega<1$) of the mean is greater than the desired
confidence level $p$ ($p<1$):
\begin{align}
  P\left(\left|\bar{x}_N-\kappa_n\right|\le \left|\omega\kappa_n\right|\right) 
     &\ge p. & \label{eqn:discussion:convergence}
\end{align}
Solving Eq.~\eqref{eqn:discussion:convergence} for $N$, we obtain:
\begin{align}
  N \ge 2 \frac{\var{\kappa_n}}{\kappa_n^2}\left(\frac{\erf^{-1}{p}}{\omega}\right)^2, 
     & \label{eqn:dicussion:nshots}
\end{align}
where $\erf^{-1}$ is the inverse error function. 

We illustrate the results obtained using Eq.~\eqref{eqn:dicussion:nshots} in
Fig.~\ref{fig:convergence}, for the scenarios discussed in
Sections~\ref{sec:discussion:dominant}--\ref{sec:discussion:marker} above.  In
all cases, we choose the desired peak convergence $\omega=20\%$, the
confidence level $p=95\%$, and do not differentiate between the true and false
contributions to the cumulant. As expected, converging higher cumulants
generally requires more shots. As already seen above for the relative peak
widths $r_n$ at a fixed number of shots (Panels d of Figs.
\ref{fig:case-a}--\ref{fig:case-d}), higher event rates $\nu_0$ improve the
convergence rate of the $2-$cumulant. This is generally not the case for the
higher cumulants, where the number of shots needed tends to \textit{increase}
with the event rate. Thus, somewhat counterintuitively, for higher cumulants
one should keep the average event rates low, to ensure faster convergence of
the cumulant.

Unsurprisingly, the level of noise has only a limited effect on the rate of
convergence, as long as the signal is dominated by the true-cumulant
contribution [Fig.~\ref{fig:convergence}(a,d), and $\kappa_3$, $\kappa_4$ in
Fig.~\ref{fig:convergence}(c)]. For the regimes where the false, noise-induced
contributions dominate [Fig.~\ref{fig:convergence}(b)], an increase in the noise
level accelerates convergence to the (contaminated) result. This behaviour is
also expected: the mean of the cumulant distributions $\kappa_n$, which appears
in the denominator of Eq.~\eqref{eqn:dicussion:nshots}, is typically more
sensitive to the noise than the variance $\var{\kappa_n}$ in the numerator.

\section{Conclusions and perspective\label{sec:conclusions}}

In this work, we have introduced the statistical analysis of cumulant
mapping\cite{Frasinski22a} in an imperfect experimental setup, where external
noise sources introduce spurious, ``false'' correlations, and therefore
``pollutes'' the intrinsic cumulant signal between fragments. Even at low noise levels,
characteristic of a well-designed laboratory apparatus and laser systems
(ca.~$1\%$), false-cumulant contribution can contaminate the signal of
interest, especially in a precision experiment. Minor fragmentation channels
appearing on a partially-correlated background
(Section~\ref{sec:discussion:minor}) are particularly affected. Coincidence
detection is to be recommended in such situations. On the other hand, for
prominent channels cumulant mapping offers significant advantages even in the
presence of moderate external noise (Section~\ref{sec:discussion:dominant}). A
particularly interesting scenario, for which cumulant mapping is eminently
suitable, is the situation where \textit{one} of the fragments is
background-free (Section~\ref{sec:discussion:marker}). There, very high average
event rates are possible, while keeping false-cumulant contamination under
control. 

We demonstrate that the convergence rate of higher cumulants with the number of
shots has a different dependence on the average event rate, compared to the
$2-$cumulant (the covariance). For the covariance, higher event rates are
always beneficial for the convergence. High event rates typically slow down the
convergence of higher cumulants.

Our analysis shows that the external noise is an important factor for the
cumulant-mapping technique, which needs to be carefully considered in planning
the experiment and in the analysis of the results. At the very least, we
recommend that the linear relationship between the cumulant signal and the
average event rate should always be verified in an experimental measurement,
e.g. by varying the target density or laser power. Deviations from linearity
are strongly indicative of the severe false-cumulant contamination. For
highly-variable light sources, binning techniques are a popular approach for
dealing with the external noise\cite{li2022:comphys, dingel2022:scirep}. There,
the characteristics of the process must be carefully considered when choosing
the bin sizes. As we demonstrate, even $1\%$ bins may lead to false-cumulant
contamination in unfavorable cases.

Although we have tried to consider some representative measurement scenarios,
they obviously cannot exhaust the full richness of this problem. The general
expressions we provide
(\cref{eqn:results:kappa2,eqn:results:kappa3,eqn:results:kappa4,eqn:results:varkappa2,eqn:results:varkappa3,eqn:results:varkappa4})
can be used in many additional situations. For still more complicated cases, we
include analytical tools\cite{supplementary}, which can be adapted to the
desired scenario. 

\section*{Acknowledgment}
J.M. gratefully acknowledges funding from the European Research Council (ERC)
under the European Union’s Horizon 2020 research and innovation programme
within a Consolidator Grant (CoG Agreement 101003142) and from the German
Research Foundation (DFG) within a Heisenberg professorship.

\end{document}